\documentstyle[aasms4]{article}

\def\etal{\hbox{\it et al.}}

\begin{document}

\def\about{$\sim$}

\title{X-Ray Emission from Rotating Elliptical Galaxies}

\author{\sl Patricia C. Hanlan and Joel N. Bregman}
\affil{Department of Astronomy, University of Michigan, Ann Arbor, MI 48109-1090\\
jbregman@umich.edu}
\authoraddr{Department of Astronomy, University of Michigan Ann Arbor, MI 48109-1090}

\begin{abstract}

The slow inward flow of the hot gas in elliptical galaxy cooling flows
is nearly impossible to detect directly due to instrumental limitations.
However, in rotating galaxies, if the inflowing gas conserves angular 
momentum, it will eventually form a disk.  The X-ray signature of this
phenomenon is a flattening of the X-ray isophotes in the inner 1-10 kpc region.
This effect is observable, so we have searched for it in X-ray observations 
of six rotating and non-rotating early-type galaxies, obtained 
mainly with the \textit{ROSAT} PSPC and HRI imagers.

The ellipticities of the X-ray emission never increase toward the
central region, nor are the X-ray ellipticities significantly greater
than the ellipticities for the optical stellar emission.
Central ellipticities in excess of 0.5 were expected in rotating 
ellipticals whereas values of 0-0.2 are measured.
The failure to detect the expected signature requires a modification to
the standard cooling flow picture, possibly including partial galactic winds,
rapid mass drop-out, or turbulent redistribution of angular momentum.

\end{abstract}



\section{Introduction}
 
Elliptical galaxies had long been thought of as purely stellar systems
until observations using the {\it{EINSTEIN Observatory}} (Gianconni \etal\ 1979)
revealed significant X-ray emission from these objects (Forman \etal\ 1979).
The emission has been attributed to a hot interstellar medium because 
of the brightness of the emission as well as the softness of the spectrum. 
The gaseous component of the X-ray emission is dominated by the stellar 
component in only the most X-ray faint objects (Canizares, Fabbiano, and
Trinchieri 1987; Irwin and Sarazin 1998).  
The comparatively faint X-ray brightness of individual galaxies, relative to 
galaxy clusters, made observations of extended objects difficult and only six individual elliptical galaxies had {\it{EINSTEIN Observatory}} high resolution images that contained more than 150 photons
(Fabbiano, Kim and Trinchieri, 1992).  The {\it{ROSAT}} 
observatory was launched with instruments on board that were more
sensitive and of lower background, than those aboard the 
{\it{EINSTEIN Observatory}} and thus it helps us gain greater insight 
into the hot interstellar medium in elliptical galaxies.

  In the steady-state cooling flow model, the source of the gas is stellar
mass loss, which collides with and shocks gas lost by other stars, resulting
in a medium at a temperature equivalent to the velocity dispersion of the stars
(5-10x$10^6$K; Thomas \etal\ 1986; Loewenstein and Mathews 1987; Sarazin and 
White 1988; Vedder \etal\ 1988; David \etal\ 1991).  The cooling time of this 
gas is much shorter than a Hubble time, and because the radiative loss rate 
increases inward, the inner regions lose pressure support more rapidly. This 
effect leads to a slow inward flow, typically only $\sim$ 5-50 km sec$^{-1}$
(e.g., Thomas 1986; Loewenstein and Mathews 1987; Vedder \etal\ 
1988; Sarazin and Ashe, 1989), inward flow of hot gaseous material.

Most steady-state, cooling flow models do not include large scale rotation
of the cooling system.  Though optically luminous elliptical galaxies do not, in general, rely 
on rotation to support the stellar system, many galaxies have 
significant rotational velocities (Franx \etal\ 1989, Davies and Birkinshaw, 1988).  We expected that as cooling gas flows inward, conserving specific angular momentum, the 
energy stored in rotation increases, so rotational support of the gas 
eventually would become significant.  Subsequently, the hot gas flows more 
rapidly along the rotation axis than in the rotational plane, which 
produces a flattening of the cooling material.  On the basis of this suggestion, we began an observing program to search for the flattening of the X-ray isophotes toward the inner parts of elliptical galaxies.

In a series of papers, detailed models of rotating cooling flows have been developed by Kley and Mathews (1995), and Brighenti and Mathews (1996, 1997).  Kley and Mathews (1995) examined the behavior of cooling flows for slowly rotating spherical elliptical galaxies.  They found that as the gas flows inward, rotational support increases, eventually preventing the gas from further inward flow.  This leads to the formation of a cooled disk, which is associated with a flattening of the X-ray isophotes.  In addition, the X-ray luminosity is significantly lower than models without rotation, which avoids the need for distributed mass drop-out assumed in other models.

This initial effort by Kley and Mathews (1995) was improved upon by Brighenti and Mathews (1996), who used more realistic models for elliptical galaxies.  These more realistic models use oblate ellipsoidal stellar density distributions with differing amounts of flattening and rotation.  They calculate the time evolution of the hot gas distribution for massive elliptical galaxies of type E0-E4.  They show that there is a significant flattening of the X-ray isophotes, from nearly spherical in the outer parts to high degrees of flattening near the location of the cooling disk (a flattening of approximately 5:1 near the disk in model E2;025).  The size of the disk depends upon the details of the models, but for all rotating models considered, the disk radius is in the range 2-15 kpc at the present epoch.
In their most recent paper, Brighenti and Mathews (1997) examined the evolution of cooling flows in less massive elliptical galaxies, where rotational support of the stellar system is relatively more important.  For these models, the X-ray emission from the hot gas in rotating ellipticals is low ($10^{38}$ erg s$^{-1}$) and would be difficult to observe.  The observations presented here should be described more accurately by the models for massive elliptical galaxies (Brighenti and Mathews 1996).

The detection of this flattening in the X-ray isophotes is not feasible with observations from the {\it{EINSTEIN Observatory}}.  The Imaging Proportional Counter (IPC) had a resolution of $1^{\prime}$, which corresponds to a linear size of 6 kpc for a galaxy at a distance of 20 Mpc.  This is too poor to resolve the disk region in most of the models.  Adequate angular resolution was available with the High Resolution Imager (HRI; $5\arcsec$ resolution), but only six HRI observations of individual elliptical galaxies with more than 150 source photons (Fabbiano, \etal\  1992), and nearly all of these galaxies lack significant projected rotation.

The {\it{ROSAT}} observatory instrumentation is well-suited for the detection of the signatures of rotating cooling flows.  The Position Sensitive Proportional Counter (PSPC) is three times more 
sensitive than the {\it{EINSTEIN}} IPC, and with twice the spatial resolution ($25\arcsec$).  It is difficult to detect structure on the same scale as the instrumental resolution, but our tests indicate that structure at least twice the instrumental resolution can be studied reliably.  Therefore, for the galaxies in our study, the PSPC will define the X-ray properties of the outer region, generally beyond the cooling disk.
The High Resolution Imager is three times less sensitive than the PSPC, but 
the spatial resolution is five times better ($5\arcsec$ at the image center), and the instrumental background is far lower than the
{\it{EINSTEIN}} HRI.  This resolution corresponds to a linear size of 0.5 kpc for a galaxy at 20 Mpc, so with suitably long observations, the region of disk formation can be resolved for the rotating models of Brighenti and Mathews (1996).

In this paper, we report on our effort to use {\it{ROSAT}} PSPC and HRI observations to search for the signatures of rotating cooling flows.  
We obtained both PSPC and HRI data for six nearby optically luminous elliptical galaxies.  The primary goal was to determine the degree of the flattening in the X-ray isophotes from the outer to the inner region of the galaxy.
Also, this sample of objects includes galaxies with both large and small 
rotation velocities, so we should be able to detect a correlation of rotation velocity with isophotal flattening in the inner regions of these galaxies, if present.

\section{Observations and Initial Reduction}

To search for the X-ray signature of rotating cooling flows, we need 
to obtain the isophotal shape of the X-ray emission in a sample of galaxies  
that contains rotating systems and a control set of
non-rotating systems.  The optical morphology and rotation parameters for 
our sample can be found in Table 1.  The optical diameters for the major 
and minor axes are from the NASA Extragalactic Database (NED), the magnitudes,
$r_e$, and the distances are from Faber \etal\ (1989; 
using $H_o$ = 70 km s$^{-1}$ Mpc$^{-1}$), ${L_X}/{L_B}$ ($L_X$ is for the 0.5 - 2.0 keV band) is from Brown and Bregman (1999),
while the rotational data come from Franx, Illingworth and Heckman, 
(1989; FIH), van der Marel, (1991; V), and the compliation of Roberts \etal\  
(1991; R).

For an observation to be useful for morphological analysis, there must
be at least approximately 300 photons from the source and the galaxy must be resolved.
This number limits the sample to nearby galaxies which are both extended and
X-ray bright.  The galaxies in our sample have optical sizes of a few 
arcminutes (see Table 1) and optical effective radii of about 0.5-1'.
The optical optical core radius, which is useful for comparison to an
X-ray core radius, is an order of magnitude smaller than the effective
radius, 3-10'' for the galaxies in this sample.  {\it{ROSAT}} PSPC observations do not 
resolve any X-ray emission within 30 arcseconds, because of the 
instrumental point spread function, so the regions resolved in the PSPC 
observations are generally beyond 10 optical core radii.  The 
resolution of the HRI lets us examine the shape of the galaxy as 
small as $5\arcsec$ to $10\arcsec$, but due to lower sensitivity and higher 
background, does not detect diffuse emission beyond $30\arcsec$ for 
most of our observations.  The observations of both instruments are needed 
to determine the ellipticities over a range of radii for a comparison to the 
predictions of Brighenti and Mathews (1996). A summary of the X-ray observations 
can be found in Table 2, which is a list of 
all of the nearby galaxies with available low and high resolution X-ray data
as well as published data pertaining to the rotation of the stellar system.  
All but two of the observations are from {\it{ROSAT}}; the HRI image of NGC 4472
and the IPC image of NGC 4374 are from the {\it{EINSTEIN}} observatory, the 
second because a good {\it{ROSAT}} image was not taken during the lifetime of the PSPC.

 To create the highest sensitivity HRI images, we used a subset of the photon 
event list of each observation.  Of the 15 HRI energy 
bins, only the channels that contain photons from the galaxy were selected 
for our usable dataset. These were, generally, channels 2 - 6 of the 15 HRI 
channels.  To maximize the signal to noise of the observation, the value of 
the S/N for a source 10\% as bright as the model background level was 
calculated and then time bins with larger and larger background levels
were added to the calculation until the S/N reached a maximum.  All time bins 
with background levels higher than those which created the maximum S/N were 
omitted from the usable data set (Pildis \etal\  1995).  The edited data 
files were blocked to create $1\arcsec \rm{x} 1\arcsec$ pixels and a 
background was determined before forming a small image of 128 x 
128 pixels for further analysis.

 The {\it{ROSAT}} PSPC images were reduced in three steps.  First, 
as with the HRI data, time intervals with high backgrounds were removed until a 
maximum S/N was achieved.  An examination of the  spectra of the central sources 
indicated that there were few source photons below energies of 0.4 keV, so an 
energy band of $0.42 \rm{keV} - 2.0 \rm{keV}$ was chosen to continue the 
analysis. Finally, the images were flattened using the program CAST\_EXP 
(Snowden \etal\ 1994; bands 3-7 used) which creates an accurate flat field by using the aspect and 
event rate information from the observation as well as the specific energy band 
and time filters.  The resulting PSPC images were blocked to create $4\arcsec 
\rm{x} 4\arcsec$ pixels and the background was determined before creating a small 
image of usually 128 x 128 pixels.  The {\it{EINSTEIN}} IPC data of NGC 4374 is an 
image file and does not contain time and energy information for each photon event, so 
all photons were used, a background was determined, and a smaller image was created.  

The PSPC images, because of their wider field of view and greater
sensitivity, often contain point sources not related to the central galaxy.  
These point sources have been edited and replaced with a background.  
This was accomplished by placing photon events randomly over the area in question 
so that the average pixel count in the edited region would match the 
previously determined background count for the observation.  When only 
the galaxy in question was left in the middle of the data frame, the
initial data reduction was considered completed.

\section{Analysis }

A morphological analysis of the data was carried out, following the 
processing of the data into blocked image files.  One of the analysis goals is
to determine a representative ellipticity for both the inner and 
outer regions of the six galaxies.  Using the STSDAS package
in IRAF, one can fit ellipses directly to the images. An advantage to this
technique is the ability to map large or complicated changes in isophotal shape;
we will use this technique for the PSPC data of NGC 1395. However, there are
two disadvantages to fitting one elliptical isophote at a time.  First,
there are large errors due to the small photon count in most X-ray images and 
secondly, the true shape of the galaxy is modified by the instrumental Point
Spread Function (PSF).  To perform a more accurate analysis, we will fit
an elliptical ``beta'' model to the data, taking into account the instrumental
resolution and the Poisson statistics of the pixel values.

 Since most X-ray images include $10^2 - 10^{3.5}$ photons, one would like to 
consider as many photons as possible when trying to make statements
about the morphology of the sources that we see.  We perform this 
fit the whole image (128 x 128 pixels) to a model surface 
brightness function:
$$S(x,y)= {S_{0} \over {(1+{1\over{a^2}}(x^2 + {y^2\over{(1-\epsilon)^2}}))^\alpha}} + S_{b}$$
$$ x=x^{\prime} cos\theta + y^{\prime} sin\theta$$
$$ y=y^{\prime} cos\theta - x^{\prime} sin\theta$$

\vspace{3pt}
\noindent
where $S_{0}$ is the central intensity, $a$ is the core radius along the 
major axis, $\epsilon$ is the ellipticity of the isophotes, $S_{b}$ is a 
flat random background and $\alpha = 3\beta -0.5$.  This function is a 
standard elliptical beta model used to fit the surface brightness profiles of elliptical 
galaxies (Cavaliere and Fusco-Femiano, 1976), which lead to a single value
for the ellipticity per fit.  The rotation of the 
distribution through a position angle $\theta$ is included
in the calculation of  $x$ and $y$ from the unrotated coordinates 
$x^{\prime}$ and $y^{\prime}$.  Before any model surface brightness 
distribution is compared to data, it is convolved with the instrumental
Point Spread Function (PSF) of the observations being modeled using a Fast 
Fourier Transform (FFT).  The PSF is approximated by a two-dimensional
Gaussian of the form $e^{{-r^2\over{2 \sigma^2}}}$, where
the width of the PSF is dependent on both the instrument and the off-axis angle 
of the target object (Hasinger \etal\  1994).

 The best fit to the surface brightness is achieved by minimizing a
statistic that is analogous to $\chi^2$, but takes into account the Poisson 
nature of the photon statistics. This maximum likelihood statistic, $\psi$, is
$$\psi = - \sum_x \sum_y D_{xy} ln(S(x,y)) + \sum_i \sum_y S(x,y) $$

\vspace{3pt}
\noindent
where  $D_{xy}$ is the image array of the X-ray data, and  $S(x,y)$ 
is the model, which depends on the parameters $S_{0}, a, 
\epsilon, \theta, \alpha$, and $S_{b}$.

To find the best-fit surface brightness distribution for any given image, a 
downhill simplex method was used to optimize the multiparameter function 
$\psi$. This optimization routine can be used for any number of parameters.  
When first trying to converge to an answer, the parameter $S_{b}$,
which was established using IRAF/PROS during the data reduction process,
was kept constant.
If the best fit core radius, $a$, is smaller than the 
half width of the instrumental PSF, then it is fixed to a value smaller than 
the PSF.  All images that were minimized were 128 x 128 pixels, with the 
exception of the PSPC image of NGC 4472, which was 512 x 512.  The larger 
data array greatly slows the calculations, because of the FFTs performed 
during each evaluation of $\psi$.  For each $ROSAT$ image, the simplex 
routine was used several times with different initial guesses at the correct 
parameter set to ensure that the algorithm consistently
converged to the correct answer.

Note that the center of the surface brightness distribution is not a 
floating parameter in the optimization routine.  The minimized imaged must be 
chosen such that the center of the galaxy is the center pixel in the 
data array.  The galaxy centers are determined using an iterative process.
First, the galaxy centers are estimated by eye using the IRAF/PROS image 
display and a minimized image is created.  The optimization code is run on 
this image, creating a rough surface brightness model for the galaxy.  To 
ensure that the center pixel of the image is truly the center of the galaxy,
the fitted surface brightness model is subtracted from the smoothed image.
If the center of the image is offset from the center of the model, the 
residual image created by subtracting the model from the data 
will be asymmetric.  Since the central emission from the galaxy is dominated 
by a PSF that is several pixels wide, the initial estimates of the galaxy 
center were usually sufficient.

 As a consistency check for our optimization method, the technique of
simulated annealing (Goffe \etal\  1994) was also used on some 
of our data.  This method for finding the minimum of a function 
randomly samples the phase 
space created by the parameters of the function to be minimized.  The 
method, because of the large sampling of phase space, is very robust and 
does an excellent job of finding the global extrema of a function.  The 
main drawback of this method is that it is slow, requiring about an order of 
magnitude more computing time to perform a single optimization than the downhill 
simplex code.  For the cases in which simulated annealing was used, 
the solution was consistent with the downhill simplex solution.

To calculate the errors in the best-fit parameters found by our optimization 
codes, a Monte Carlo method was adopted.  For each best-fit surface brightness 
model, $S_{best}(S_0,a,\epsilon,\alpha,\theta,S_{bck})$,
we have simulated a series of {\it{ROSAT}} observations of this optimal
distribution  convolved with a known PSF. The simulated 
observations have signal-to-noise ratios, background noise, and total photon 
counts which match the {\it{ROSAT}} observation currently being modeled.  
Poisson deviates were taken from the convolved best-fit model surface 
brightness, $S_{best}$, until the number of ``photons''
 placed in the simulated image matched the photon count of the 
{\it{ROSAT}} dataset in question. 

 For each $ROSAT$ image and corresponding best-fit surface brightness 
distribution model, 50 to 100 of these simulated datasets were created and 
the parameters of their surface brightness distribution estimated by 
optimizing $\psi$.  Since all of the simulated datasets came from the same 
surface brightness distribution, the variation in the calculated best fit 
parameters is an estimate of the accuracy of the analysis method.   
The standard deviation of the distribution 
of the best fit parameters are used as error bars for the best 
fit parameters of the real $ROSAT$ observation.

As stated previously, the $ROSAT$ PSPC image of NGC 1395 cannot be 
well fit by a single ellipticity and so the optimization code was not 
used. Instead, elliptical contours were fit to the galaxy using the 
task ELLIPSE, in the STSDAS package in IRAF.  This task 
calculates the ellipticity and position angle, at some semi-major
axis, of elliptical contours given an initial position for the galaxy 
center, an initial guess for the semi-major axis, ellipticity, and 
position angle of the first contour. Due to the Poisson nature of the 
unsmoothed data, most of the pixels have zero values.  The routine 
ELLIPSE relies on the assumption that the distribution to be fit is one that 
changes fairly smoothly.  Consequently, this routine is unsuccessful 
at fitting unsmoothed ROSAT data.  However, when the data are smoothed 
with a Gaussian comparable to the instrumental PSF of a given image, 
the pixel-to-pixel variation is less severe and ellipses can be fit 
successfully.

Since a smoothed image is being used, one needs to consider how this will 
affect the results returned by the ellipse-fitting routine.   An elliptical 
object that, on the sky, has a semi-major axis $a$ and semi-minor axis $b$ 
is convolved with a Gaussian smoothing function that reflects the
instrumental resolution aboard {\it{ROSAT}}.  The observed axes of an 
elliptical isophote will be approximately the sum, in quadrature,
of the true axes and the half-width of the smoothing function; for the
semi-major axis,
$$ a_{obs}^2 \simeq a_{true}^2 + r_{PSF}^2 .$$
Consequently, the ellipticity, $\epsilon$, will be reduced by the
instrumental resolution.  The most severely affected case arises for a small, 
flattened ellipse; an ellipse observed to have an ellipticity of 0.5 and a 
semi-major axis $a=10\arcsec\ $ will have a 12\% error in the estimate of 
the ellipticity, {\it{i.e.}}, the true ellipticity of the observed object 
would be 0.563.  The above expression is applicable only to noiseless data, 
so a more reliable assessment of the ellipse fitting procedure and the 
accuracy of the error bars is called for in the analysis of the ROSAT data.  
Again, we have performed Monte Carlo simulations for this purpose, using 
simulated images of the PSPC data for NGC 1395 and analyzing these simulated 
data with the task ELLIPSE. The ELLIPSE task fits isophotes at radii from 4 to 40 arcseconds, but we only considered the isophotes outside of the instrumental PSF, trying to avoid one of the main pitfalls of this method of analysis.
The uncertainty in the fit to NGC 1395 was determined from an ensemble of simulated data sets, for an ellipticity of 0.1 and processed with ELLIPSE.

\section{Results }

 The processed X-ray observations are presented in Figures \ref{n1395cont} 
to \ref{p5322cont}.  Each processed model image was smoothed with a 
Gaussian that matches the instrumental PSF of the observation.  For the 
PSPC data of NGC 1404, the PSF has a half-width of $37^{{\prime}{\prime}}$  
because the image of NGC 1404 is not in the center of the PSPC frame, 
but $10^{\prime}$ off-axis. Contour maps of the X-ray observations, 
both PSPC and HRI, are plotted with contours $3\sigma$, 
$9\sigma$, $27\sigma$, ... etc. above the local background.  For the 
images with shallow surface brightness profiles, an extra contour that 
helps define the center of the X-ray emission was added.
 
 Model surface brightness distributions were fit to the data using the 
optimization code described above.  The best-fit parameters, along 
with their respective errors, are given in Table 3. 
Since the values of $\psi$ in phase space are symmetric about the 
$\epsilon = 0$ axis, and are periodic along the $\theta$ axis,
the best-fit parameters are all shown with positive ellipticities and with 
position angles having a range of $0^{\circ} - 180^{\circ}$.  
The position angles are set with $0^{\circ}$ pointing North and 
positive angle increasing counterclockwise. 
For most 
simulations, the core radius, $a$, often unresolved, was kept fixed and was 
only changed in increments of 1.0 pixels in separate runs of the optimization code 
to determine the minimum value of $\psi$ that could be found in parameter 
space.  For the simulations of the HRI observation of NGC 1404 and NGC 4552, 
the errors were calculated using an optimization code that permitted $a$ 
to vary.  The estimate of the errors in $a$, $\pm 1$ pixel ($1\arcsec$), come 
from these two sets of simulations; the values quoted in Table 3 are 
in arcseconds.

The shape of the X-ray emission as a function of radius is what we would 
like to compare to the predictions for rotating cooling flows. 
Table 4 shows the derived ellipticity 
of the HRI and PSPC images as well as optical data at two radii ($10\arcsec$ 
and $30\arcsec$) that correspond to the X-ray data.  The optical data comes 
from work by Djorgovski (1985) and all errors are in brackets behind
each number.  Following is a list of how the optical and X-ray shape of
each galaxy are related to one another.

{\bf{NGC 1395}} -- This galaxy has one of the largest velocities of 
rotation among the elliptical galaxies detectable by ROSAT, 93 km sec$^{-1}$, and optical observations 
(Malin and Carter, 1983, Forbes and Thomas, 1992) of NGC 1395 show substructure in the form of 
shells.  The PSPC data are not regular, i.e. a satisfactory best-fit parameter set could not be 
calculated, possibly in response to the event that created the shells, 
and we did not attempt to fit a single ellipticity 
to the observation.  Notice the large error on the position angle of the 
modeled PSPC surface brightness.  The roundness of the galaxy and 
the low X-ray photon count in the HRI observation made the determination of a position 
angle very difficult.  Given the uncertainty, both the PSPC ellipticities are 
constrained with the HRI ellipticities, which are consistent with the shape of the 
optical ellipticities.
  
{\bf{NGC 1404}} -- This galaxy is near the center of the Fornax cluster 
(within $10\arcmin$ of the central galaxy, NGC 1399) and has a rotation 
velocity of 90 km sec$^{-1}$.    
The HRI ellipticity is not different than the PSPC ellipticity to within 
the error. 

{\bf{NGC 4374}} -- M84 is a member of the Virgo cluster and has a very low
line of sight rotation velocity (8 km sec$^{-1}$).  NGC 4374 is also a radio galaxy 
(3C272.1) with jets that have  position angles of $-5^{\circ}$ and 
$170^{\circ}$  (Laing and Bridle, 1987) and are linear for $40\arcsec$
before they are engulfed by a larger diffuse radio emission.  The core of 
this galaxy also contains dust lanes (Goudfrooij, 1994), situated at a 
position angle of $80^{\circ}$. The HRI ellipticities 
are not significantly rounder than the IPC ellipticities. 
The position angle of the HRI emission puts it in areas that avoid the radio
emission, as seen by Laing and Bridle (1987).
The optical and X-ray ellipticities agree with one another.

{\bf{NGC 4472}} -- The X-ray bright galaxy NGC 4472 (M49) is a luminous Virgo 
cluster member with a moderate rotation velocity of 27 km sec$^{-1}$.  The PSPC 
ellipticities are flatter than the HRI ellipticities and are flatter
than the optical isophotes of the galaxy at the same radii. 
The HRI and optical ellipticities are similar.  

{\bf{NGC 4552}} -- M89, like NGC 1395, is a shell galaxy (Malin, 1979), 
but with a low rotation velocity (15 km sec$^{-1}$).  The PSPC and HRI ellipticities 
are the same within calculated errors, but the X-ray emission is significantly flatter than the optical emission at all radii; this is the only galaxy for 
which this occurs.  We note that this is the only S0 galaxy in the sample.

{\bf{NGC 5322}} -- This X-ray faint galaxy has a rotation velocity of
40 km sec$^{-1}$.  All of the ellipticities are the same to within the uncertainties.
The HRI observation has only 360 photons from the source in the 128 x 
128 arcsecond field and this paucity of counts is reflected in the large 
error in the optimized ellipticity.

\section{Discussion and Conclusions }

A primary goal is to examine whether the X-ray emission from rotating
ellipticals provides insight into the flow of the hot gas, so we review the
qualitative predictions of the models and compare them to our observations.
The predictions for the model depend upon a variety of important issues,
such as whether rotation is responsible for the optical ellipticity, which
we examine first. \ In low luminosity galaxies, rotation can account for the
observed ellipticity of the optical isophotes, while in the high luminosity
galaxies, rotation is insufficient, so an anisotropic velocity dispersion is
required (Davies \etal\ 1983; Binney and Tremaine 1987). \ These galaxies
have an average optical luminosity that is a bit higher than galaxies whose
flattening is due to rotation, but by examining the properties of the
galaxies in more detail, more definitive statements about rotational support
can be made.

The two fastest rotators are NGC\ 1404 and NGC 1395, so they have the
greatest probability of having rotation contribute to their flattening. \
The galaxy NGC 1404 appears to have an optical ellipticity that is close to
the value expected from rotation. For elliptical galaxies whose isodensity
surfaces are coaxial oblate spheroids (Binney and Tremaine 1987), the
observed ratio of the mean rotational velocity to the line-of-sight velocity
dispersion, v/$\sigma $ = 0.37-0.40 (Davies \etal\ 1983; Franx, Illingworth,
and Heckman 1989), would lead to an expected ellipticity of 0.08-0.09. Since
the observed optical ellipticity is 0.12-0.14, much of it could be due to
rotation. However, in the central region $15\arcsec$ of the galaxy, the rotational
velocity decreases sharply, so the mean v/$\sigma $ in this region is 0.20,
which would produce an ellipticity of 0.02. As the observed optical
ellipticity in the center is significantly greater ($\epsilon $ = 0.14),
the flattening must be due largely to an anisotropic dispersion.

The other rapidly rotating galaxy, NGC 1395, a system of similar optical
luminosity has a rotation curve that rises to a value of 93 km sec$^{-1}$ at a
radius of $15\arcsec$, and beyond this radius, the value of v/$\sigma $ is
0.36-0.45, depending on which weighting of v or $\sigma $ is used (Davies et
al. 1983; Franx, Illingworth, and Heckman 1989); within $15\arcsec$ the mean v/$%
\sigma $ is 0.15-0.20. The ellipticities associated with these values of v/$%
\sigma $ for the rotational flattening model (above) are $\epsilon $=
0.01-0.02 at r = $10\arcsec$ and $\epsilon $= 0.08-0.11 for r $>$
$15\arcsec$. However, the optical ellipticity is observed to have a mean value of $%
\epsilon $= 0.15 for r $\leq$ $10\arcsec$ and $\epsilon $= 0.18 for r $>$
$15\arcsec$. Evidently, an anisotropic velocity dispersion is responsible for the
flattening in the center and for about half of the flattening in the outer
regions.

For the three galaxies with $\epsilon \approx 0.10$, NGC 4374, NGC 4472,
and NGC 4552, the observed  v/$\sigma$ is low, 0.03-0.11, 
whereas rotational flattening would require v/$\sigma
\approx $ 0.4, so the flattening in these systems must be dominated by the
anisotropic velocity dispersion. \ Similarly, for NGC\ 5322, v/$\sigma $%
=0.13, but for rotation to produce the observed optical ellipticity of $\epsilon
\approx 0.4$, a value of v/$\sigma $ = 1 would be needed. \ To conclude,
velocity anisotropy rather than rotation is the likely cause of the observed
optical flattening for the galaxies in our sample. \ This indicates that the
most appropriate models are those for the higher mass galaxies (Brighenti
and Mathews 1996), and it is for systems like these that they developed
their models.

The basic expectations from the models were that the X-ray isophotes in the
inner 1-10 kpc region would be flatter than either the stellar distribution
or the X-ray isophotes in the outer part of the galaxy (beyond 10 kpc). \
These predictions are not supported by our data. \ For the most rapid
rotators, NGC 1395 and NGC 1404, the X-ray isophotes are no more elliptical
than the optical isophotes, and the ellipticities for the PSPC and HRI data
are indistinguishable. \ One would expect these rapid rotators to have their
rotation axis close to the plane of the sky, so the degree of flattening
should be similar to that shown by Brighenti and Mathews (1996),
approximately $\epsilon $=0.5-0.8 (model E2;025), while the observed
ellipticities are typically 0.1 (Table 4). \ In the other four more slowly
rotating galaxies, none show a significant increase in the ellipticity of
the HRI relative to the PSPC.

A difference between the HRI\ and PSPC\ ellipticities occurs in only one
case, NGC 4472, but it is in the sense that the ellipticity becomes greater
at radii beyond $3^{\prime}$. \ Irwin and Sarazin (1996) 
have discussed this flattening
in the outer part of NGC 4472 and they attribute it to the interaction of
the galaxy with the environment of the Virgo cluster. \ Such interactions
have been noticed previously in the Virgo cluster, such as in the case of
NGC 4406 (White \etal\ 1991). \
Interactions of this type may explain why NGC 4552, also a Virgo cluster
member, has PSPC\ and HRI ellipticities slightly flatter than the optical
ellipticities.

The basic angular momentum cooling flow model appears to be ruled out, but
it is difficult to determine if one should consider entirely different
models or if modifications to this model are possible. \ There are a few
possible explanations for the absence of the flattening predicted in the
X-ray isocontours, most of which are discussed in Brighenti and Mathews
(1996). \ In order to avoid the expected flattening, it is necessary that
the gas fails to move inward to a fraction of its initial radius. \ Galactic
winds will naturally prevent infall, but unless the outward flow of the gas
is halted by a surrounding medium, the X-ray luminosity will fall far below
the observed levels. \ Partial winds are another possibility, where some gas
flows outward beyond a stagnation radius while the gas interior to that
radius flows inward. \ To assess the viability of this picture, one would
need to calculate if the predicted flattening within the central region is
consistent with the limits set by the HRI (within $10\arcsec$, or about 0.1-0.3 R$%
_{e}$ for these galaxies) while still producing the observed X-ray
luminosity. \ In support of the total or partial galactic wind picture, we
note that Davis and White (1996) find that the hot gas temperature is
significantly above the stellar velocity dispersion temperature, as might be
expected if a wind or breeze were present.  Alternatively, the relatively
high X-ray temperatures may be due to deeper potential wells due to 
extended massive dark halos.

Another possibility for avoiding flattened disks is that the gas cools
before it falls to a fraction of its original size. \ In several cooling
flow models, distributed mass drop-out is included and is attributed to the
development of thermal instabilities. \ Unfortunately, the physical basis
for this is questionable since linear thermal instability modes generally
are not unstable in this environment (Balbus 1991). \ Also, thorough angular
momentum mixing could prevent disk formation. \ This was considered for
cooling flows during the early stages of galaxy formation (Fabian and Nulsen
1994) but it might be applicable today as well. \ They suggest that
turbulence (and shocks) create an effective viscosity that mixes the angular
momentum. \ While this is not meant to be a definitive discussion of
alternative models, it shows a range of possibilities that must be
considered before we have a thorough understanding of the phenomenon. \
Several of these models make predictions that may be possible to test with
more detailed measurements of the density and temperature distribution as
well as of the metallicity of the X-ray gas, and such observations are
likely to be undertaken with the new generation of telescopes \textit{Chandra%
} and \textit{XMM}.


We would like to thank W.G. Mathews, C.S. Sarazin,  J. Irwin, D.O. Richstone, E. Schulman, and R. Pildis for their comments and insights.  Also, we would like to thank
the ROSAT team for an outstanding effort in all aspects of the observations.
During the course of this research, use was made
of the NASA/IPAC Extragalactic Database (NED) which is
operated by the Jet Propulsion Laboratory, California Institute of
Technology, under contract with the National Aeronautics and Space
Administration.  Support for this program was provided by NASA through
NAG5-3247, NAG5-1955, and NAGW-4448.

\clearpage

\begin{center}
REFERENCES
\end{center}

\def\ARAA{{\it Ann. Rev. Astron. Astrophys.}}
\def\ApJ{{\it Ap. J.}}
\def\ApJL{{\it Ap. J. Lett.}}
\def\ApJSS{{\it Ap. J. Supp. Ser.}}
\def\AandA{{\it Astron. Astrophys.}}
\def\AJ{{\it Astron. J.}}
\def\JCP{{\it J. Comp. Phys.}}
\def\MNRAS{{\it M. N. R. A. S.}}
\def\N{{\it Nature}}
\def\PASJ{{\it Publ. Astron. Soc. Jap.}}
\def\RPP{{\it Rep. Prog. Phys.}}

\begin{verse}

Balbus, S.A. 1991, {\apj}, {\bf372}, 25.

Binney, J. and Tremaine, S. 1987, {\it Galactic Dynamics }, (Princeton:
 Princeton University Press)

Brighenti, F., and Mathews, W.G. 1996, {\apj}, {\bf 470}, 747.

Brighenti, F., and Mathews, W.G. 1997, {\apj}, {\bf 490}, 592.

Brown, B.A., and Bregman, J.N. 1999, {\apj}, submitted.

Canizares, C. R., Fabbiano, G., \& Trinchieri, G. 1987, {\apj}, {\bf 312}, 503

Cavaliere and Fusco-Femiano 1976, { \aap}, {\bf 49}, 137.

Davis, D.S., and White, R.E. 1996, {\apj}, {\bf 470}, L35.

Djorgovski 1985, {\it Thesis}, University of California, Berkeley.

David, L. P., Forman, W. and Jones C. 1991, {\apj}, {\bf 369}, 121.

Davies, R. L., Efstathiou, G., Fall, S. M., Illingworth, G. and Schechter, P. L. 1983, {\apj}, {\bf 266}, 41.

Davies, R. L. and Birkinshaw, M. 1988, {\apjs}, {\bf 68}, 409.


Fabbiano, G., Kim, D.-W. and Trinchieri, G. 1992, {\apjs}, {\bf 80}, 531.

Faber, S. M., Wegner, G., Burstein, D., Davies, R. L., Dressler, A., Lynden-Bell, D., \& Terlevich, R. J. 1989, {\apjs}, {\bf 69}, 763.

Fabian, A. C. and Nulsen, P. E. J. 1994, {\mnras}, {\bf 269}, L33.

Forbes, D. A. and Thomson, R. C. 1992, {\mnras}, {\bf 254}, 723.

Forman, W., Schwarz, J., Jones, C., Liller, W. and Fabian, A. C. 1979, {\apj}, {\bf 234}, 27.

Franx, M., Illingworth, G. and Heckman, T. 1989, {\apj}, {\bf 344}, 613.

Goffe, Ferrier, and Rogers, 1994, {\it Journal of Econometrics}, {\bf 60}, 65.

Giacconi, R., Branduard, G., Briel, U., Epstein, A., Fabricant, D., Feigelson, E., Holt, S. S., Becker, R. H., Boldt, E. A. and Serlemitsos, P. J., 1979, {\apj}, {\bf 230}, 540.

Goudfrooij, P., Hansen, L., Norgaard-Nielsen, H. U., Jorgensen, H., E., de Jong, T. and and den Hoek, L. B. 1994, {\apjs}, {\bf 105}, 341.

Hasinger, G., Boese, G., Predehl, P., Turner, T. J., Yusaf, R., George, I. M., and Rhrbach, G. 1994, MPE/OGIP Calibration Memo CAL/ROS/93-015

Irwin, J. and Sarazin, C. 1996, {\apj}, {\bf 471}, 683.

Irwin, J. and Sarazin, C. 1998, {\apj}, {\bf 499}, 650.

Kley, W. and Mathews, W. G. 1995, {\apj}, {\bf 438}, 100.

Laing, R. A. and Bridle A. H. 1987, {\mnras}, {\bf 228}, 557.

Loewenstein, M. and Mathews, W. G. 1987, {\apj}, {\bf 319}, 614.

Malin, D. F. and Carter, D. 1983, {\apj}, {\bf 274}, 534.

Malin, D. F. 1979, {\it Nature}, {\bf 277}, 279.

Pildis, R. A., Bregman, J. N. and Evrard, A. E. 1995, {\apj}, {\bf 443}, 514.

Roberts, M. S., Hogg D. E., Bregman, J. N., Forman, W. R., and Jones, C. 1991, {\apjs}, {\bf 75}, 751.

Sarazin, C. L. and Ashe, G. A. 1989, {\apj}, {\bf 345}, 22.

Sarazin, C. L. and White, R. E. 1988, {\apj}, {\bf 331}, 102.

Schombert, J. M., Barsony, M. and Hanlan, P. C. 1993, {\apj}, {\bf 416}, L61.

Snowden, S. L., McCammon, D., Burrows, D. N., and Mendenhall, J. A. 1994, {\apj}, {\bf 424}, 714.

Thomas, P. A. 1986, {\mnras}, {\bf 220}, 949.


Vedder, P. W., Trester, J. J., and Canizares, C. R. 1989, {\apj}, {\bf332}, 725.

van der Marel, R. P. 1991, {\mnras}, {\bf 253}, 710.

White, D. A., Fabian, A. C.,  Forman, W. R., Jones, C., and Stern, C. 1991, {\apj}, {\bf 375}, 35.

\end{verse}

\clearpage

\begin{deluxetable} {ccccccccccc}
\tablecaption{Galaxy Optical Morphology and Rotation Parameters}
\tablehead{
\colhead {Galaxy} & 
\colhead {Type} & 
\colhead {B$_{T}^{0}$} & 
\colhead {Diameter} & 
\colhead {v$_{rotation}$} & 
\colhead {$\sigma$} & 
\colhead {r$_{e}$} & 
\colhead {d} & 
\colhead {logL$_{B}$} & 
\colhead {Log$\frac{L_{X}}{L_{B}}$} & 
\colhead {Ref.}\nl
\colhead {} &
\colhead {}  & 
\colhead {} & 
\colhead {(arcmin)} & 
\colhead {(km s$^{-1})$} & 
\colhead {(km s$^{-1})$} & 
\colhead {(arcsec)} & 
\colhead {(Mpc)} & 
\colhead {(erg s$^{-1})$} & 
\colhead {($\frac{{erg s}^{-1}}{{L}_{\odot}}$)} & 
\colhead {}
}
\startdata
NGC 1395 & E2 & 10.94 & 5.9$\times$4.5 & 93 & 258 & 45 & 28.4 & 43.42 &
30.02 & FIH\nl
NGC 1404 & E2 & 10.89 & 3.3$\times$3.0 & 90 & 225 & 27 & 20.3 & 43.15 &
30.53 & FIH\nl
NGC 4374 & E1 & 10.13 & 6.5$\times$5.6 & 8 & 290 & 54 & 19.0 & 43.40 & 30.10 &
V\nl
NGC 4472 & E1 & 9.32 & 10.2$\times$8.3 & 27 & 250 & 104 & 19.0 & 43.72 &
30.45 & FIH\nl
NGC 4552 & S01 & 10.84 & 5.1$\times$4.7 & 13 & 273 & 30 & 19.0 & 43.11 &
30.21 & R\nl
NGC 5322 & E4 & 11.09 & 5.9$\times$3.9 & 40 & 310 & 35 & 23.7 & 43.21 &
29.31 & R\nl
\enddata
\end{deluxetable}

\clearpage

\begin{deluxetable}{lllc}
\tablecaption{X-ray Observations 
\label{xobs}}
\tablehead{
\colhead {Galaxy} &
\colhead {Satellite} & 
\colhead {Instrument} & 
\colhead {Time (ksec)} 
}
\startdata
NGC 1395  &  {\it{ROSAT}}    & HRI  & 16.3 \nl
NGC 1395  &  {\it{ROSAT}}    & PSPC & 21.0 \nl
NGC 1404  &  {\it{ROSAT}}    & HRI  & 71.6 \nl
NGC 1404  &  {\it{ROSAT}}    & PSPC & 30.0 \nl
NGC 4374  &  {\it{ROSAT}}    & HRI  & 26.5 \nl
NGC 4374  &  {\it{EINSTEIN}} & IPC  & 35.0 \nl
NGC 4472  &  {\it{EINSTEIN}} & HRI  & 33.8 \nl
NGC 4472  &  {\it{ROSAT}}    & PSPC & 26.0 \nl
NGC 4552  &  {\it{ROSAT}}    & HRI  & 30.1 \nl
NGC 4552  &  {\it{ROSAT}}    & PSPC & 17.0 \nl
NGC 5322  &  {\it{ROSAT}}    & HRI  & 20.3 \nl
NGC 5322  &  {\it{ROSAT}}    & PSPC & 35.0 \nl 
\enddata
\end{deluxetable}

\clearpage


\begin{deluxetable}{llrrrrrr}
\scriptsize
\tablecaption{Best-fit Parameters
\label{bestfits}}
\tablehead{
\colhead {Galaxy} & 
\colhead {Instrument} & 
\colhead {$S_0(\sigma_{S_0})$}  & 
\colhead {$a(\sigma_a)$} & 
\colhead {$\epsilon(\sigma_{\epsilon})$} & 
\colhead {$\alpha(\sigma_{\alpha})$} & 
\colhead {$\theta(\sigma_{\theta})$} & 
\colhead {$S_b(\sigma_{S_b})$} 
}
\startdata
 NGC 1395 &  HRI &  1.98(0.40) &  1.0(1.0) &  0.08(0.08) & 0.55(0.04) & 130(60) & 0.023(0.006) \nl
 NGC 1395 & PSPC &          &              &   0.1(0.08) &            &  34(35) &        \nl 
 NGC 1404 &  HRI & 17.42(0.32) &  4.5(1.0) &  0.09(0.02) & 0.935(0.007) &   165(9) & 0.107(0.004) \nl
 NGC 1404 & PSPC &141.00(20.0) &  6.0(4.0) &  0.04(0.08) & 0.88(0.04) &  137(20) & 0.010(0.001) \nl 
 NGC 4374 &  HRI &  2.53(0.14) &  4.8(1.0) &  0.14(0.05) & 0.80(0.02) & 108(82) & 0.004(0.004) \nl
 NGC 4374 &  IPC &  2.80(0.12) & 40.0(8.0) &  0.19(0.06) & 1.23(0.06) &  77(10) & 0.002(0.003) \nl 
 NGC 4472 &  HRI &  14.18(2.7) &  4.0(1.4) &  0.08(0.06) & 1.00(0.15) & 185(76) & 0.66(0.11) \nl
 NGC 4472 & PSPC &  61.14(1.5) &  8.0(4.0) &  0.26(0.04) & 0.78(0.04) &  49(86) & 0.0006(0.0001) \nl 
 NGC 4552 &  HRI &  5.24(0.36) &  5.0(1.0) &  0.23(0.04) & 0.98(0.08) & 179(7) & 0.019(0.003) \nl
 NGC 4552 & PSPC &  4.90(0.85) & 12.0(4.0) &  0.24(0.13) & 2.34(0.33) & 182(73) & 0.0011(0.0002) \nl 
 NGC 5322 &  HRI &  0.88(0.30) &  3.0(1.0) &  0.11(0.14) & 1.14(0.12) &  48(60) & 0.016(0.001) \nl
 NGC 5322 & PSPC &  5.76(0.60) & 12.0(4.0) &  0.22(0.09) & 1.67(0.11) &  92(74) & 0.0022(0.0005) \nl 
\enddata
\end{deluxetable}

\clearpage
\begin{deluxetable}{lccccc}
\tablecaption{Ellipticity Measurments
\label{epsresults}}
\tablehead{
\colhead {Galaxy} &
\colhead {v$_{rot}$/$\sigma$} &
\colhead {$\epsilon_{opt} (10\arcsec)$} &
\colhead {$\epsilon_{opt} (30\arcsec)$} &
\colhead {$\epsilon_{HRI}$} &
\colhead {$\epsilon_{PSPC}$} 
}
\startdata
NGC 1395  & 0.36 & 0.1  (0.02) & 0.18 (0.02) & 0.08 (0.08) &  0.10 (0.1) \nl
NGC 1404  & 0.40 & 0.14 (0.02) & 0.14 (0.02) & 0.09 (0.02) &  0.04 (0.08) \nl
NGC 4374  & 0.03 & 0.2  (0.02) & 0.12 (0.02) & 0.14 (0.05) &  0.19 (0.06) \nl
NGC 4472  & 0.11 & 0.07 (0.03) & 0.15 (0.02) & 0.08 (0.06) &  0.26 (0.04) \nl
NGC 4552  & 0.05 & 0.04 (0.02) & 0.08 (0.02) & 0.23 (0.04) &  0.24 (0.13) \nl
NGC 5322  & 0.13 & 0.25 (0.02) & 0.20 (0.02) & 0.11 (0.14) &  0.22 (0.09) \nl 
\enddata
\end{deluxetable}


\clearpage 
\plotone{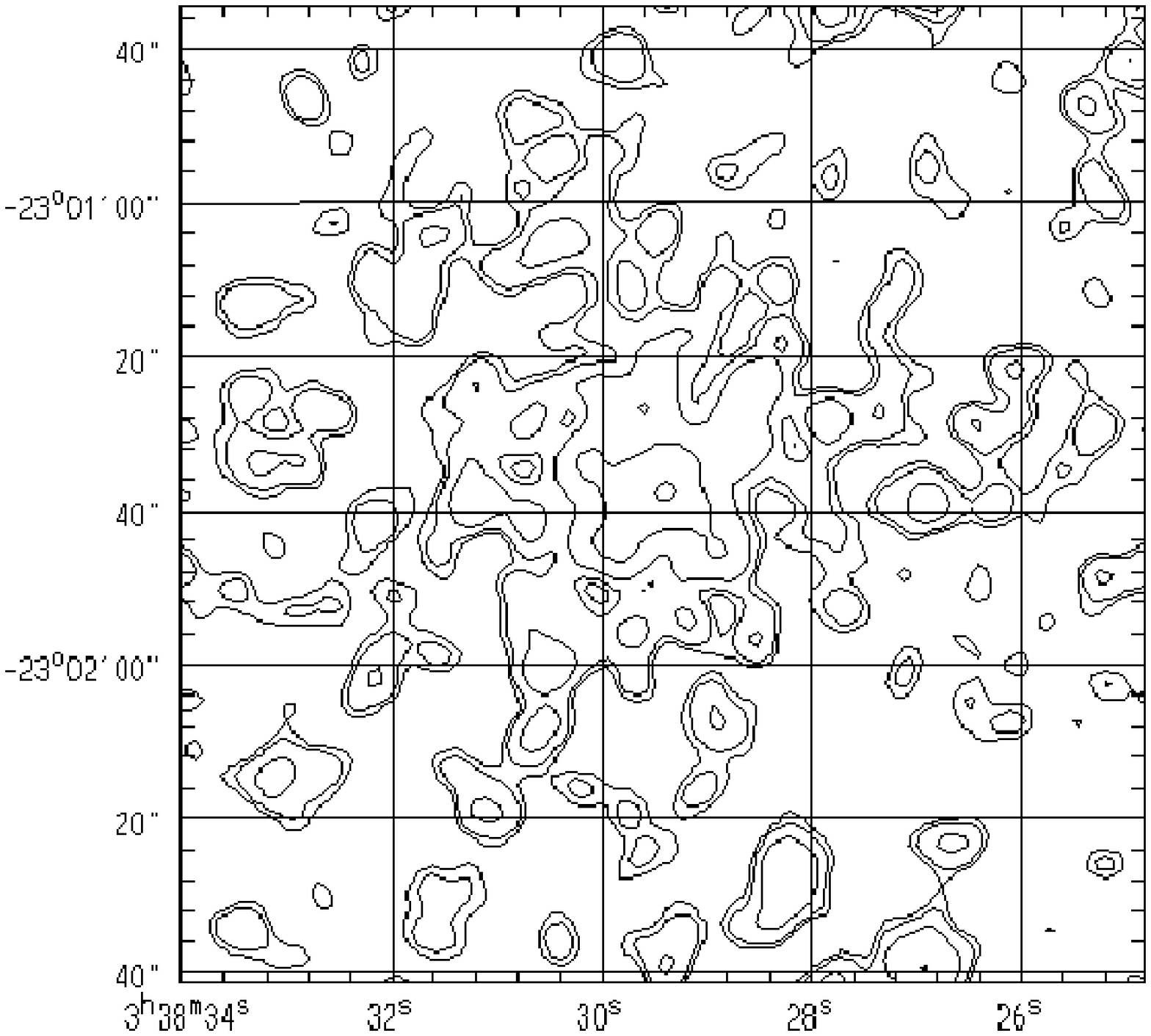}
\figcaption{ HRI contour map of NGC 1395.  The pixels are $1^{\prime\prime} 
{\rm{x}} 1^{\prime\prime}$ in size and the frame is 128 x 128 pixels.  The 
cleaned data were blocked to an instrumental beam width of $5^{\prime\prime}$.  \label{n1395cont}}

\clearpage 
\plotone{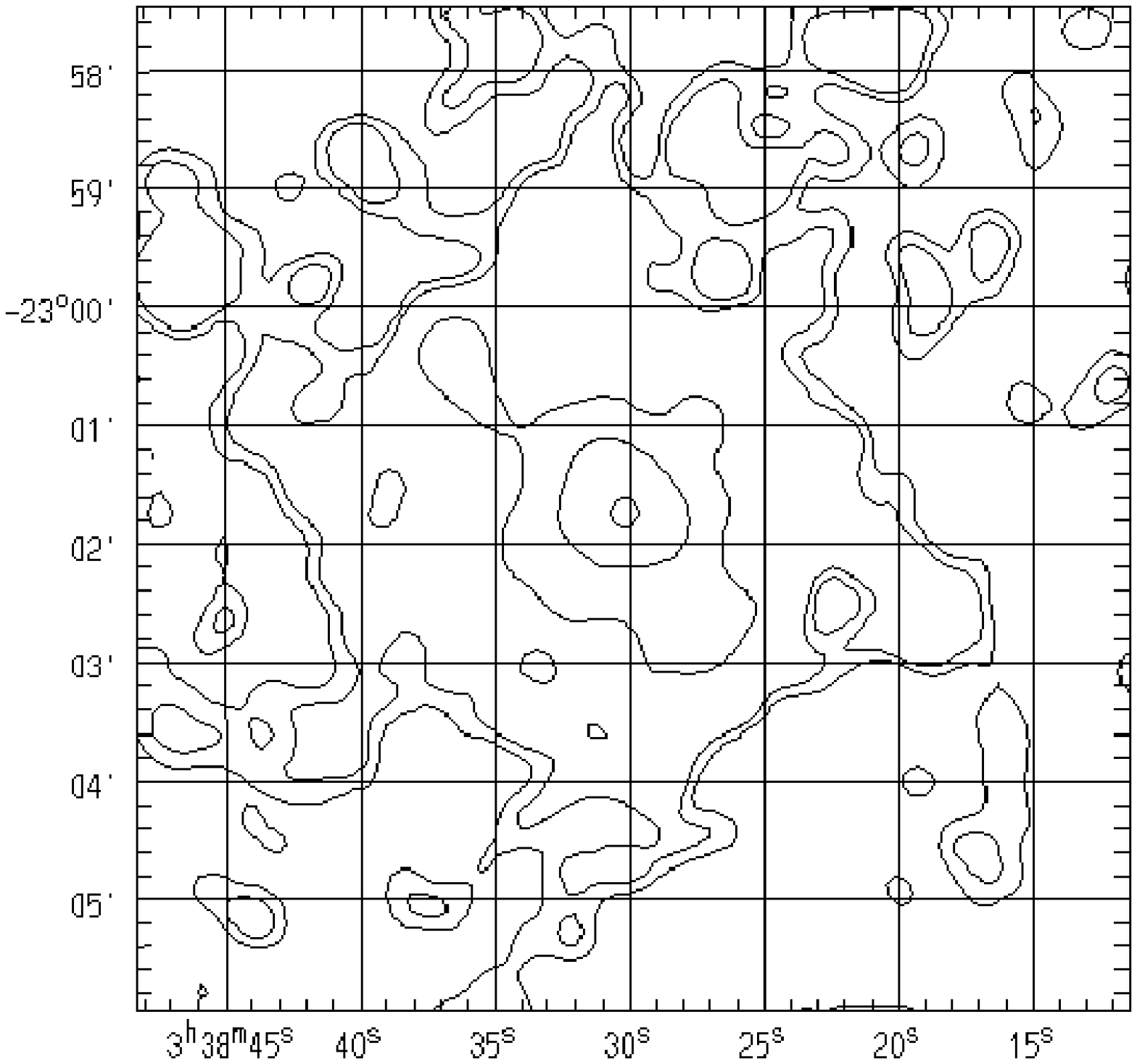}
\figcaption{PSPC contour map of NGC 1395.  The pixels are
$4^{\prime\prime} {\rm{x}} 4^{\prime\prime}$ in size and the frame is 128 x 
128 pixels.  The cleaned data were blocked  to an instrumental beam width 
of $25^{\prime\prime}$. \label{p1395cont}}

\clearpage 
\plotone{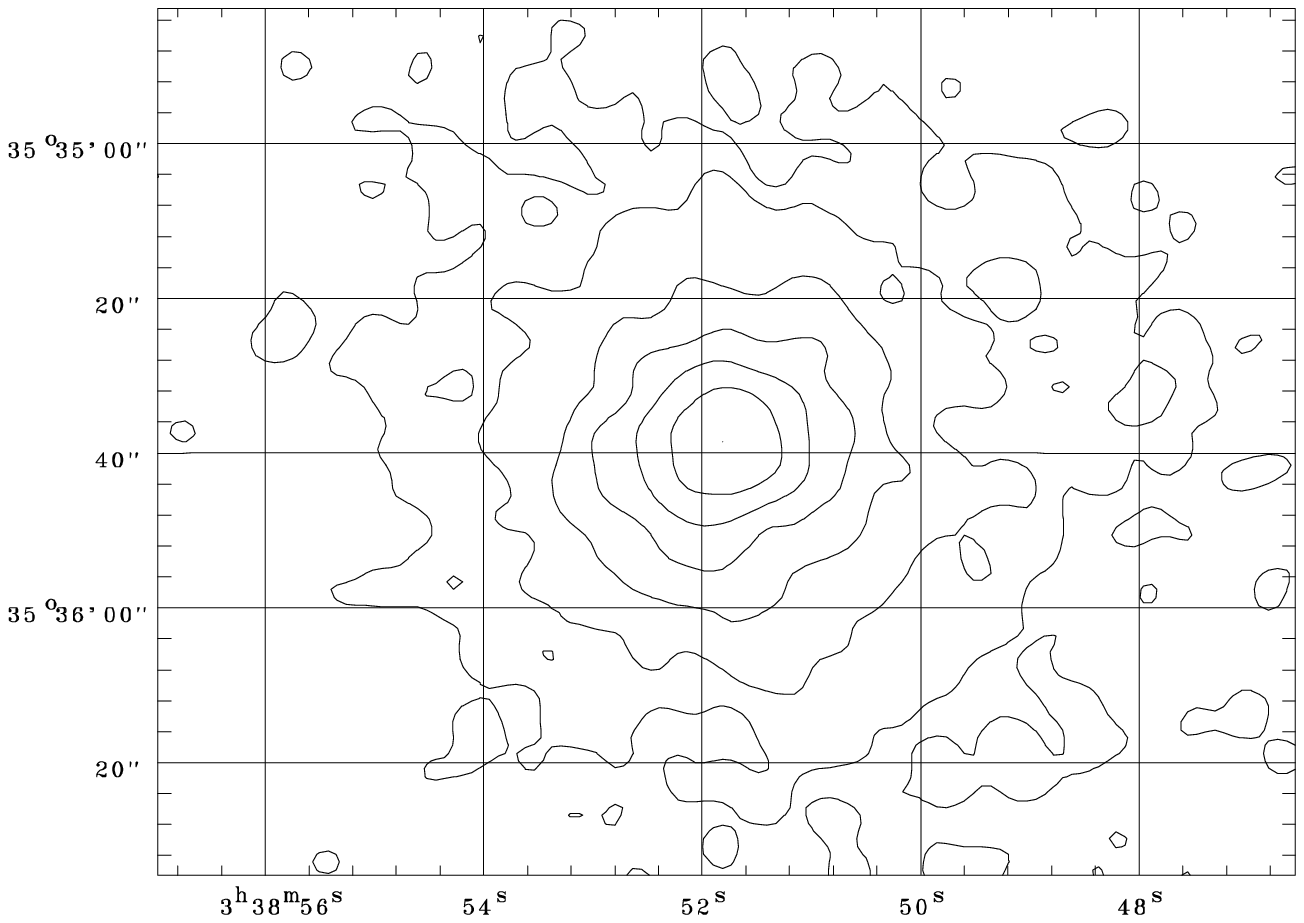}
\figcaption{HRI contour map of NGC 1404.  The pixels are
$1^{\prime\prime} {\rm{x}} 1^{\prime\prime}$ in size and the frame is 128 x 
128 pixels.  The cleaned data were blocked  to an instrumental beam width 
of $5^{\prime\prime}$. \label{n1404cont}}

\clearpage 
\plotone{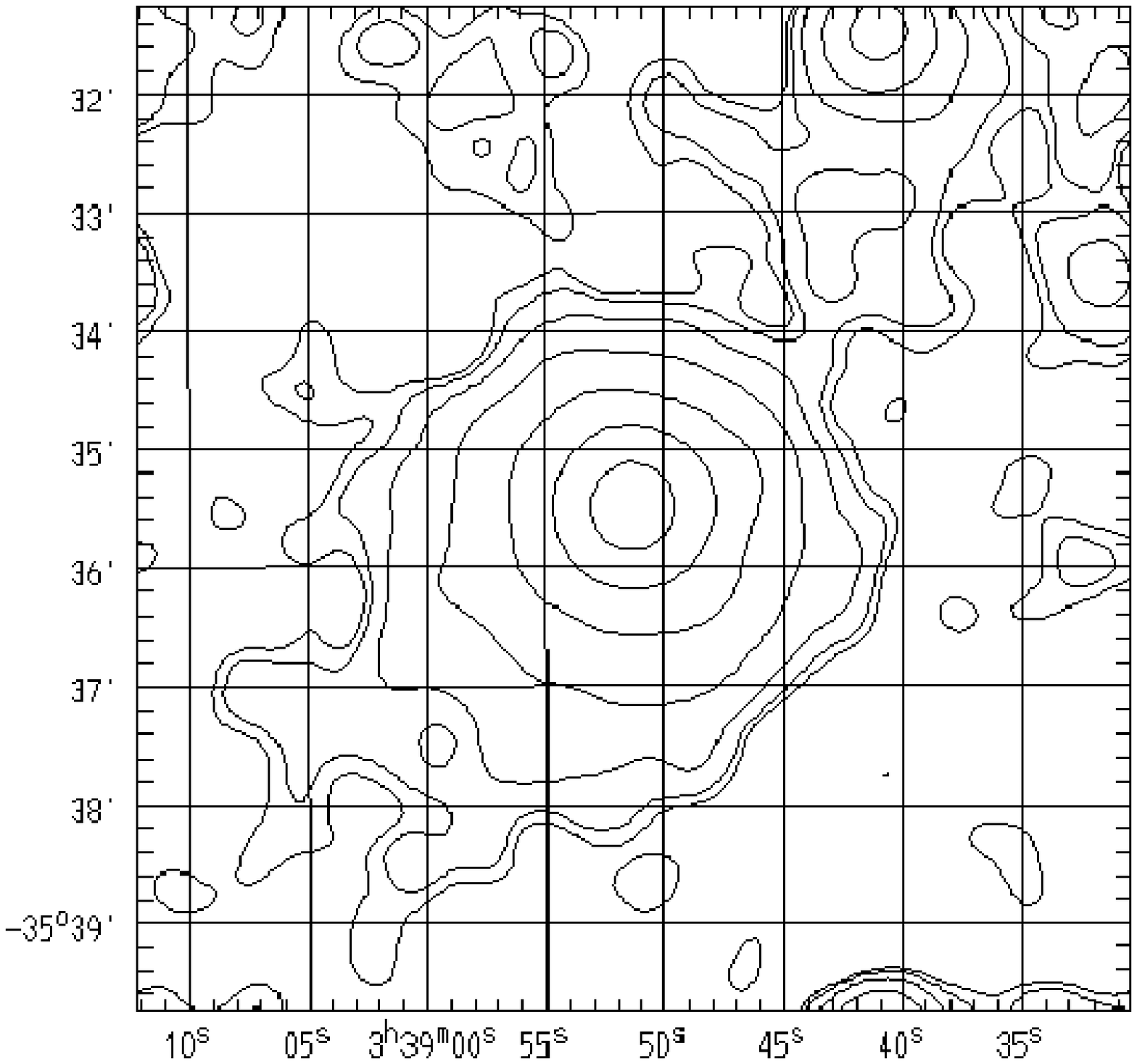}
\figcaption{PSPC contour map of NGC 1404.  The pixels are
$4^{\prime\prime} {\rm{x}} 4^{\prime\prime}$ in size and the frame is 128 x 
128 pixels.  The cleaned data were blocked  to an instrumental beam width 
of $37^{\prime\prime}$, since the image of NGC 1404 is $10^{\prime}$ of axis. 
\label{p1404cont}}

\clearpage 
\plotone{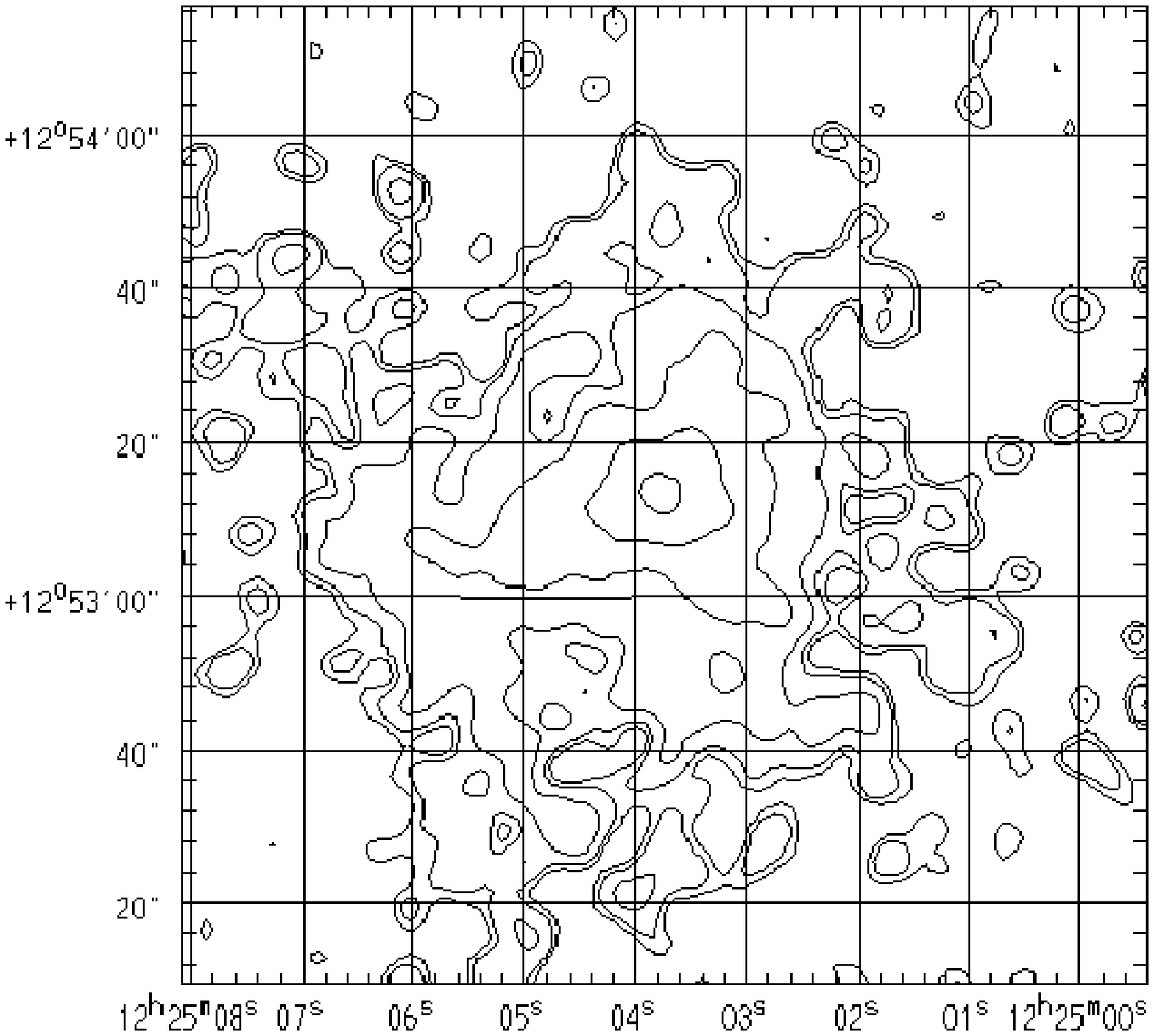}
\figcaption{HRI contour map of NGC 4374.  The pixels are
$1^{\prime\prime} {\rm{x}} 1^{\prime\prime}$ in size and the frame is 128 x 
128 pixels.  The cleaned data were blocked  to an instrumental beam width 
of $5^{\prime\prime}$. \label{n4374cont}}

\clearpage 
\plotone{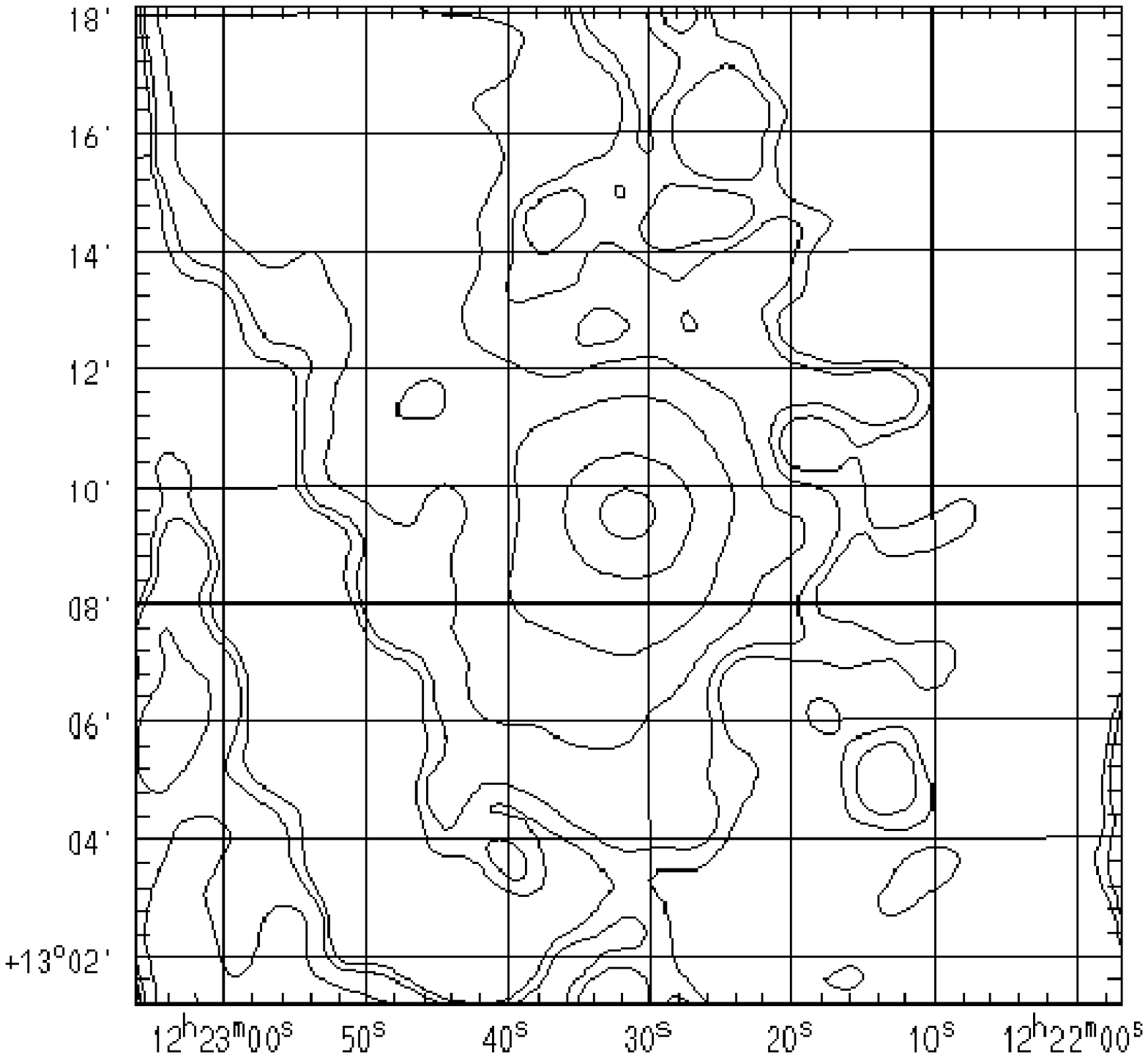}
\figcaption{ {\it{EINSTEIN}} IPC contour map of NGC 4374.  The pixels are
$8^{\prime\prime} {\rm{x}} 8^{\prime\prime}$ in size and the frame is 128 x 
128 pixels.  The cleaned data were blocked  to an instrumental beam width 
of $1^{\prime}$. \label{p4374cont}}

\clearpage 
\plotone{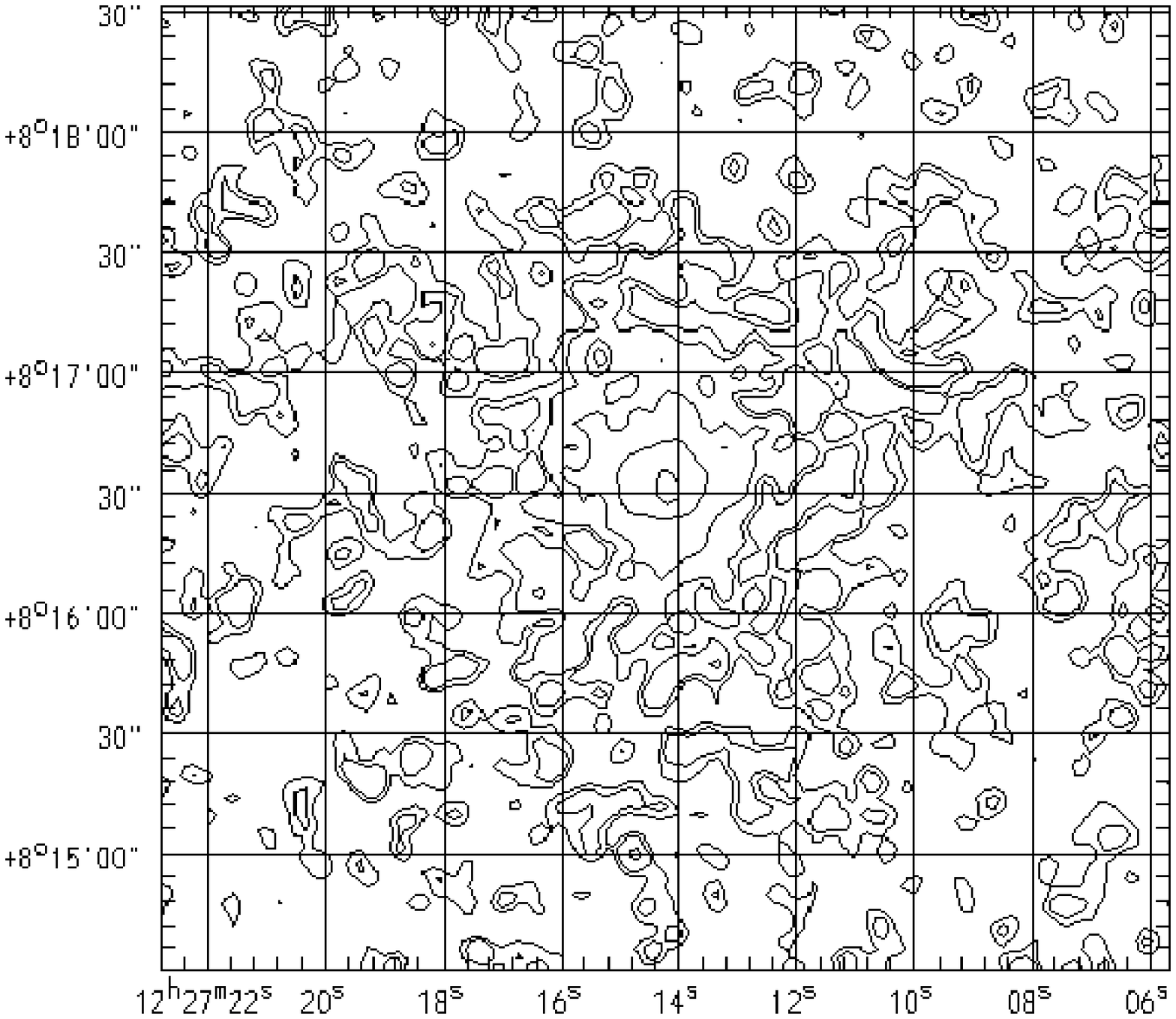}
\figcaption{ {\it{EINSTEIN}} HRI contour map of NGC 4472. The pixels are
$1^{\prime\prime} {\rm{x}} 1^{\prime\prime}$ in size and the frame is 128 x 
128 pixels.  The cleaned data were blocked  to an instrumental beam width 
of $5^{\prime\prime}$. \label{n4472cont}}

\clearpage 
\plotone{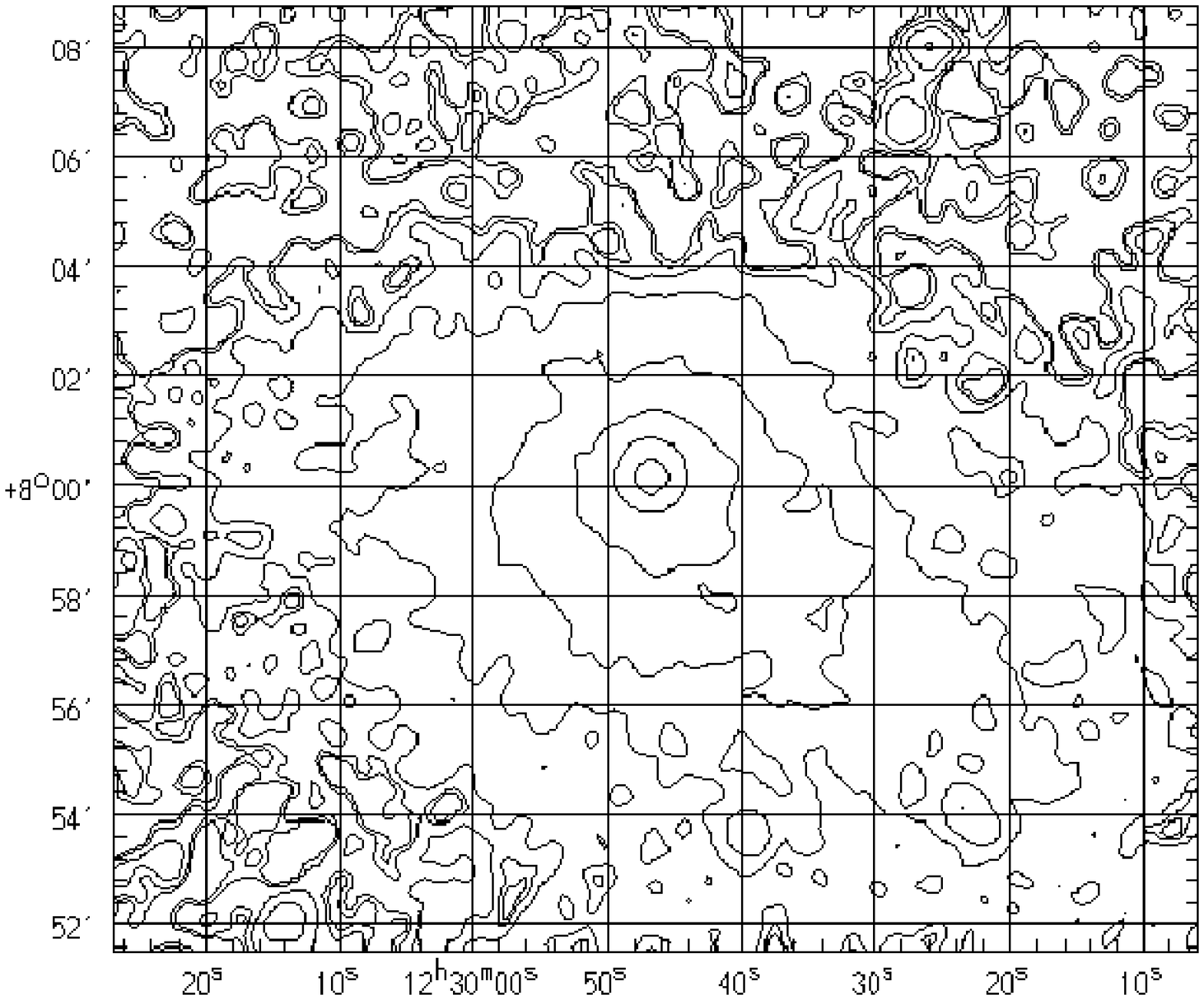}
\figcaption{PSPC contour map of NGC 4472.  The pixels are
$4^{\prime\prime} {\rm{x}} 4^{\prime\prime}$ in size and the frame is 
512 x 512 pixels.  The cleaned data were blocked  to an instrumental beam 
width of $25^{\prime\prime}$. \label{p4472cont}}

\clearpage 
\plotone{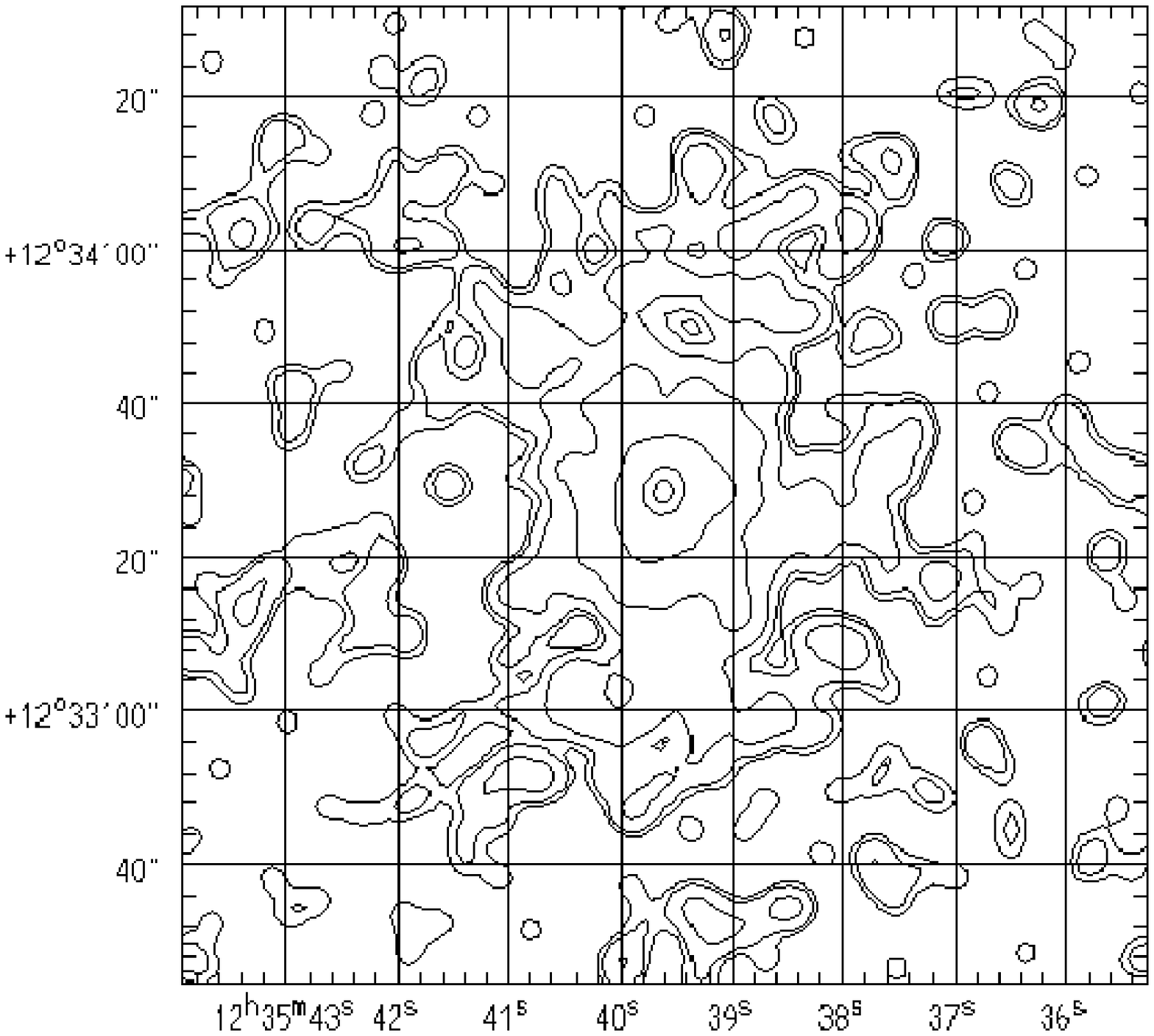}
\figcaption{ HRI contour map of NGC 4552.  The pixels are
$1^{\prime\prime} {\rm{x}} 1^{\prime\prime}$ in size and the frame is 128 x 
128 pixels.  The cleaned data were blocked  to an instrumental beam width 
of $5^{\prime\prime}$. \label{n4552cont}}

\clearpage 
\plotone{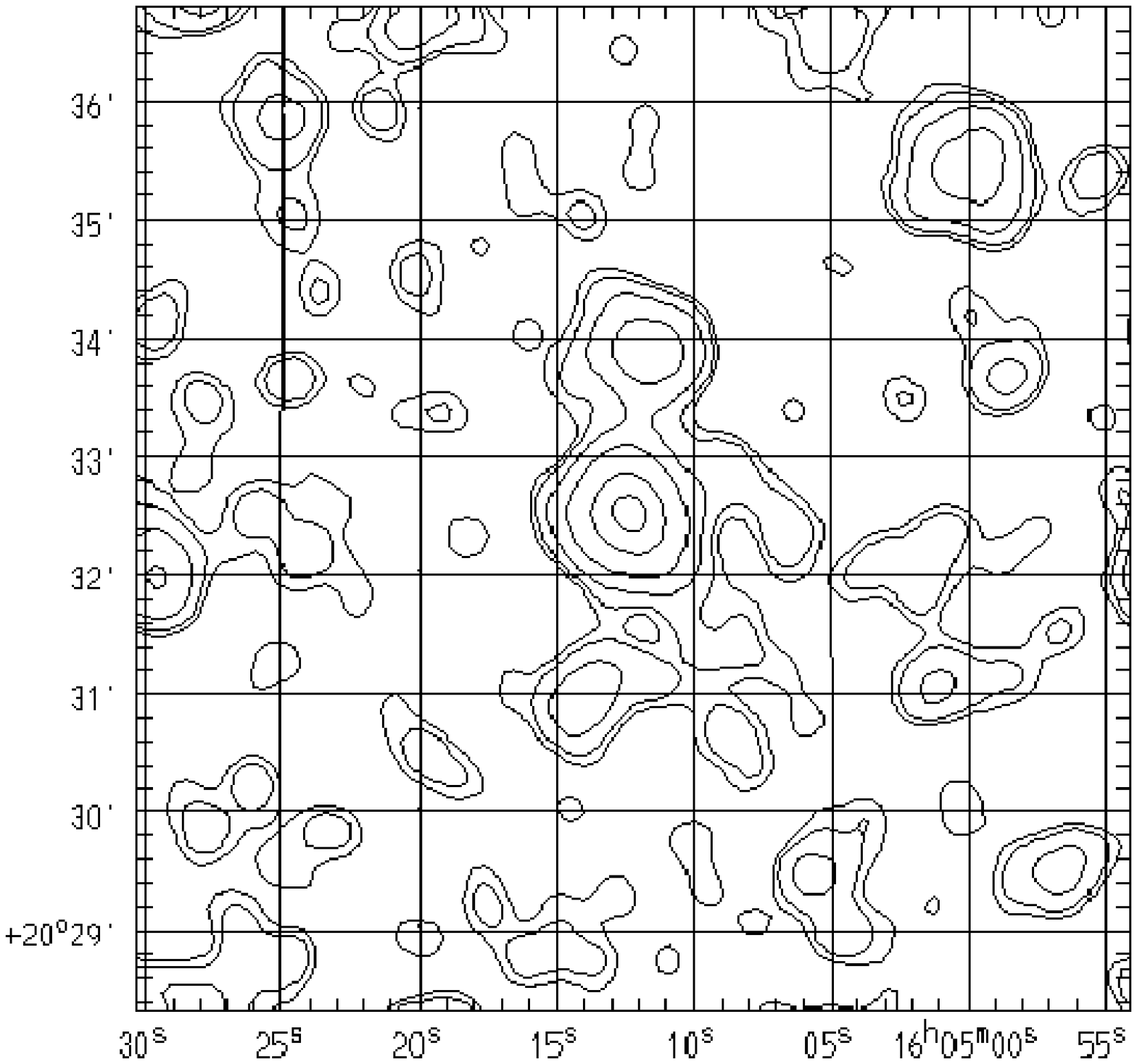}
\figcaption{PSPC contour map of NGC 4552.  The pixels are
$4^{\prime\prime} {\rm{x}} 4^{\prime\prime}$ in size and the frame is 128 x 
128 pixels.  The cleaned data were blocked  to an instrumental beam width 
of $25^{\prime\prime}$. \label{p4552cont}}

\clearpage 
\plotone{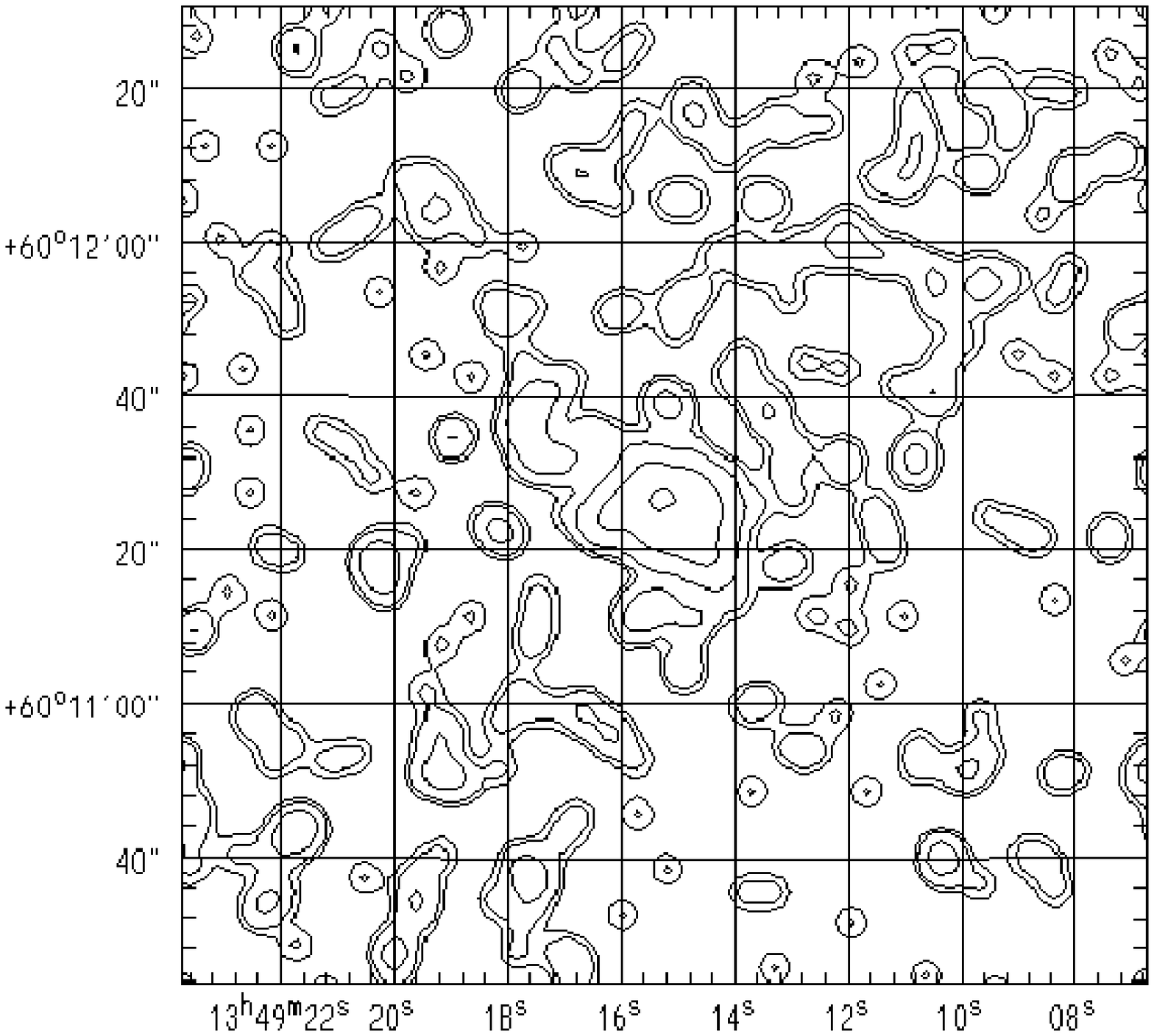}
\figcaption{HRI contour map of NGC 5322.  The pixels are
$1^{\prime\prime} {\rm{x}} 1^{\prime\prime}$ in size and the frame is 128 x 
128 pixels.  The cleaned data were blocked  to an instrumental beam width 
of $5^{\prime\prime}$. \label{n5322cont}}

\clearpage 
\plotone{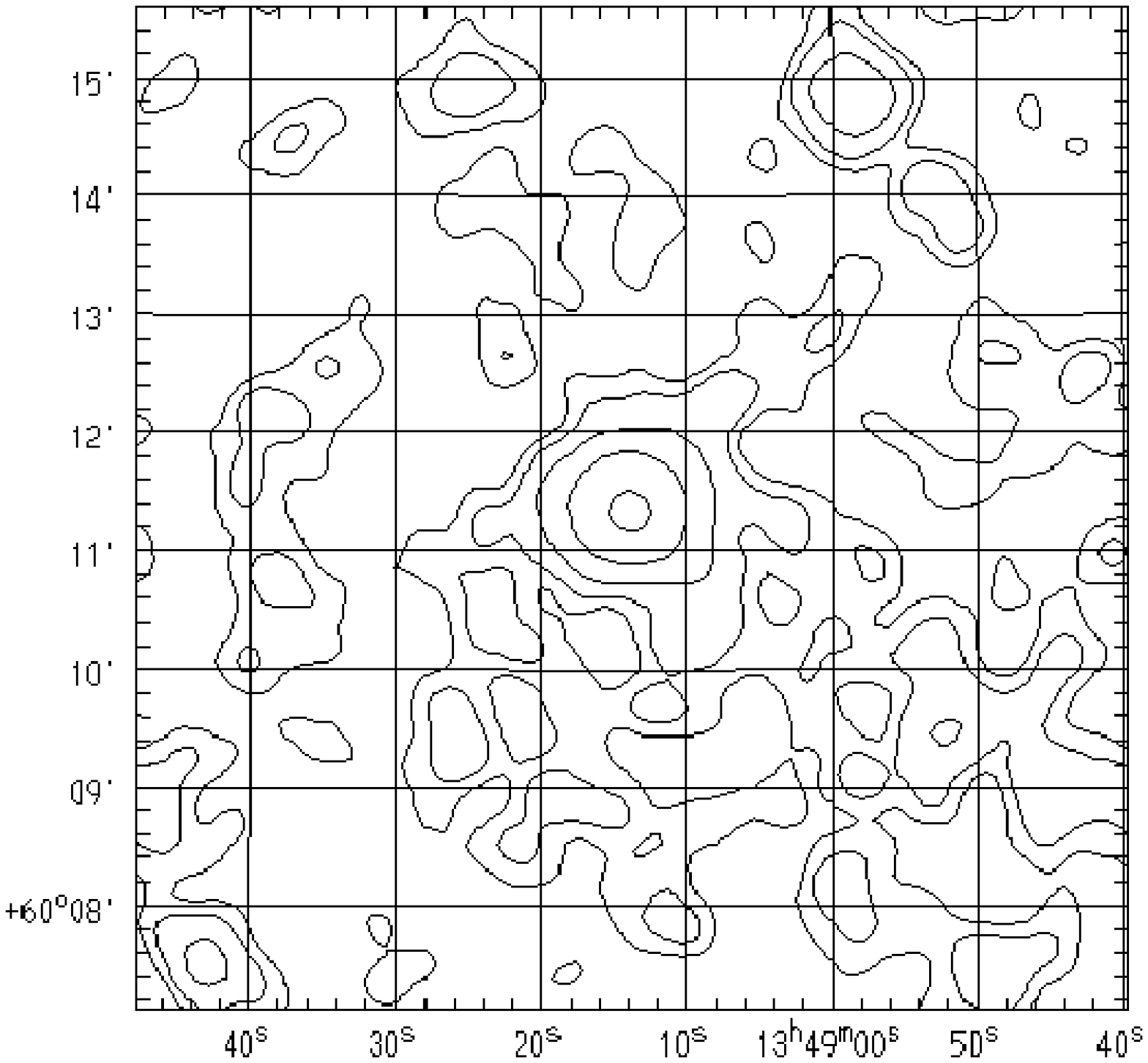}
\figcaption{PSPC contour map of NGC 5322.  The pixels are
$4^{\prime\prime} {\rm{x}} 4^{\prime\prime}$ in size and the frame is 128 x 
128 pixels.  The cleaned data were blocked  to an instrumental beam width 
of $25^{\prime\prime}$. \label{p5322cont}}

\clearpage
\plotone{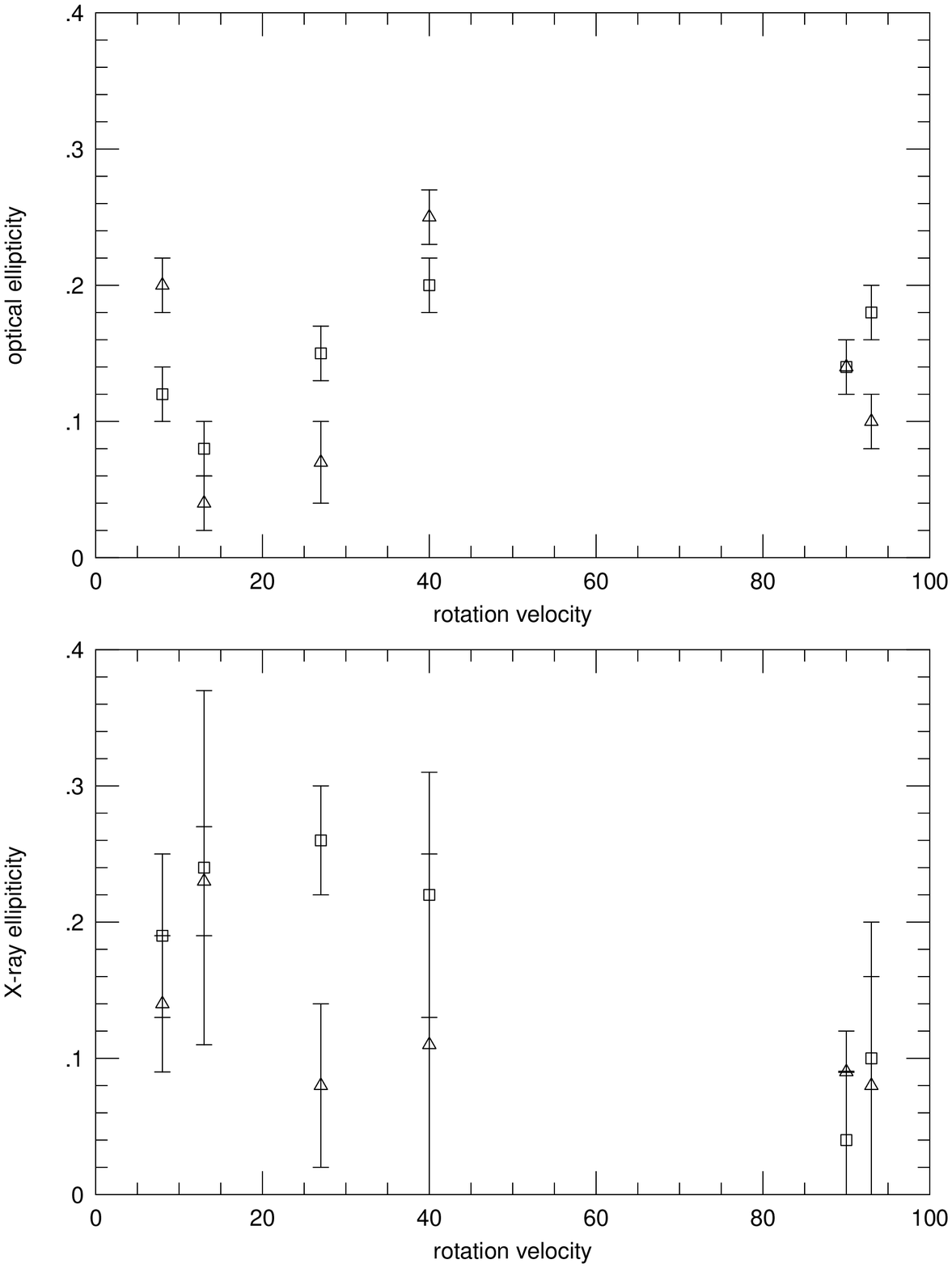}
\figcaption{ Optical and X-ray ellipticities plotted against
rotation velocity in km sec$^{-1}$ for our sample.  For each galaxy, an interior
(square) and exterior (triangle) ellipticity is plotted: in the optical,
from radii of 10 and 30 arcseconds (Djorgovski, 1985), and in the X-ray,
from the HRI and PSPC data.  \label{vrot}}

\end{document}